\documentclass[twocolumn,showpacs,aps,prl,superscriptaddress]{revtex4}

\usepackage{graphicx}
\usepackage{dcolumn}
\usepackage{amsmath}
\usepackage{xspace}

\newcommand{\ed}[1]{#1}

\newcommand{\re}{\ensuremath{\mathop{\rm Re}}}
\newcommand{\im}{\ensuremath{\mathop{\rm Im}}}
\newcommand{\sign}{\ensuremath{\mathop{\rm sgn}}}

\newcommand{\ocalA}{\ensuremath{\overline{\calA}}\xspace}

\newcommand{\Bd}{\ensuremath{\B_d}\xspace}
\newcommand{\tauB}{\ensuremath{\tau_{\Bz}}\xspace}
\newcommand{\dM}{\deltamd}
\newcommand{\G}{\ensuremath{\Gamma_d}\xspace}
\newcommand{\dG}{\ensuremath{\Delta\G}\xspace}
\newcommand{\dGoG}{\ensuremath{\dG/\G}\xspace}
\newcommand{\qop}{\ensuremath{q/p}\xspace}
\newcommand{\poq}{\ensuremath{p/q}\xspace}
\newcommand{\absqop}{\ensuremath{|\qop|}\xspace}
\newcommand{\lCP}{\ensuremath{ \lambda_{\CP} }\xspace}

\newcommand{\rmtag}{\ensuremath{{\rm tag}}}
\newcommand{\rmrec}{\ensuremath{{\rm rec}}}

\newcommand{\rmflav}{\ensuremath{{\rm flav}}}

\newcommand{\Brec}{\ensuremath{\B_{\rmrec}}\xspace}
\newcommand{\Btag}{\ensuremath{\B_{\rmtag}}\xspace}
\newcommand{\BCP}{\ensuremath{\B_{\CP}}\xspace}
\newcommand{\Bflav}{\ensuremath{\B_{\rmflav}}\xspace}

\newcommand{\z}{\ensuremath{{\mathsf z}}\xspace}
\newcommand{\imZ}{\ensuremath{\im\z}\xspace}
\newcommand{\reZ}{\ensuremath{\re\z}\xspace}
\newcommand{\sgndGoG}{\ensuremath{\sign(\relCP)\dGoG}\xspace}
\newcommand{\relCP}{\ensuremath{ \re \lCP}\xspace}
\newcommand{\imlCP}{\ensuremath{ \im \lCP}\xspace}
\newcommand{\imlambcpflat}{\ensuremath{ \im \lCP
  /|\lCP|}\xspace}
\newcommand{\reZparflat}{\ensuremath{\left( \relCP /
  |\lCP| \right) \reZ }\xspace}

\RequirePackage{xspace}

\usepackage{relsize}
\def\babar{\mbox{\slshape B\kern-0.1em{\smaller A}\kern-0.1em
    B\kern-0.1em{\smaller A\kern-0.2em R}}}

\def\pip   {\ensuremath{\pi^+}\xspace}
\def\pim   {\ensuremath{\pi^-}\xspace}

\def\Kbar  {\kern 0.2em\overline{\kern -0.2em K}{}\xspace}

\def\Kz    {\ensuremath{K^0}\xspace}
\def\Kzb   {\ensuremath{\Kbar^0}\xspace}
\def\KzKzb {\ensuremath{\Kz \kern -0.16em \Kzb}\xspace}
\def\Kp    {\ensuremath{K^+}\xspace}
\def\Km    {\ensuremath{K^-}\xspace}

\def\KpKm  {\ensuremath{\Kp \kern -0.16em \Km}\xspace}
\def\KS    {\ensuremath{K^0_{\scriptscriptstyle S}}\xspace} 
\def\KL    {\ensuremath{K^0_{\scriptscriptstyle L}}\xspace} 
\def\Kstarz  {\ensuremath{K^{*0}}\xspace}

\def\Dbar    {\kern 0.2em\overline{\kern -0.2em D}{}\xspace}

\def\Dz      {\ensuremath{D^0}\xspace}
\def\Dzb     {\ensuremath{\Dbar^0}\xspace}
\def\DzDzb   {\ensuremath{\Dz {\kern -0.16em \Dzb}}\xspace}
\def\Dp      {\ensuremath{D^+}\xspace}
\def\Dm      {\ensuremath{D^-}\xspace}

\def\DpDm    {\ensuremath{\Dp {\kern -0.16em \Dm}}\xspace}

\def\B       {\ensuremath{B}\xspace}
\def\Bbar    {\kern 0.18em\overline{\kern -0.18em B}{}\xspace}
\def\Bb      {\ensuremath{\Bbar}\xspace}
 
\def\Bz      {\ensuremath{B^0}\xspace}
\def\Bzb     {\ensuremath{\Bbar^0}\xspace}
\def\BzBzb   {\ensuremath{\Bz {\kern -0.16em \Bzb}}\xspace}
\def\Bu      {\ensuremath{B^+}\xspace}
\def\Bub     {\ensuremath{B^-}\xspace}

\def\BpBm    {\ensuremath{\Bu {\kern -0.16em \Bub}}\xspace}

\def\BorBbar    {\kern 0.18em\optbar{\kern -0.18em B}{}\xspace}
\def\DorDbar    {\kern 0.18em\optbar{\kern -0.18em D}{}\xspace}
\def\KorKbar    {\kern 0.18em\optbar{\kern -0.18em K}{}\xspace}

\def\jpsi     {\ensuremath{{J\mskip -3mu/\mskip -2mu\psi\mskip 2mu}}\xspace}

\mathchardef\Upsilon="7107
\def\Y#1S{\ensuremath{\Upsilon{(#1S)}}\xspace}

\def\FourS {\Y4S}

\mathchardef\Deltares="7101
\mathchardef\Xi="7104
\mathchardef\Lambda="7103
\mathchardef\Sigma="7106
\mathchardef\Omega="710A

\def\Deltabar{\kern 0.25em\overline{\kern -0.25em \Deltares}{}\xspace}
\def\Lbar{\kern 0.2em\overline{\kern -0.2em\Lambda\kern 0.05em}\kern-0.05em{}\xspace}
\def\Sigbar{\kern 0.2em\overline{\kern -0.2em \Sigma}{}\xspace}
\def\Xibar{\kern 0.2em\overline{\kern -0.2em \Xi}{}\xspace}
\def\Obar{\kern 0.2em\overline{\kern -0.2em \Omega}{}\xspace}
\def\Nbar{\kern 0.2em\overline{\kern -0.2em N}{}\xspace}
\def\Xb{\kern 0.2em\overline{\kern -0.2em X}{}\xspace}

\newcommand{\tev}{\ensuremath{\mathrm{\,Te\kern -0.1em V}}\xspace}
\newcommand{\gev}{\ensuremath{\mathrm{\,Ge\kern -0.1em V}}\xspace}
\newcommand{\mev}{\ensuremath{\mathrm{\,Me\kern -0.1em V}}\xspace}
\newcommand{\kev}{\ensuremath{\mathrm{\,ke\kern -0.1em V}}\xspace}
\newcommand{\ev}{\ensuremath{\mathrm{\,e\kern -0.1em V}}\xspace}
\newcommand{\gevc}{\ensuremath{{\mathrm{\,Ge\kern -0.1em V\!/}c}}\xspace}
\newcommand{\mevc}{\ensuremath{{\mathrm{\,Me\kern -0.1em V\!/}c}}\xspace}
\newcommand{\gevcc}{\ensuremath{{\mathrm{\,Ge\kern -0.1em V\!/}c^2}}\xspace}
\newcommand{\mevcc}{\ensuremath{{\mathrm{\,Me\kern -0.1em V\!/}c^2}}\xspace}

\def\mus  {\ensuremath{\rm \,\mus}\xspace}

\def\ps   {\ensuremath{\rm \,ps}\xspace}

\def\mus        {\ensuremath{\,\mu{\rm s}}\xspace}    
\def\ps         {\ensuremath{{\rm \,ps}}\xspace}  

\def\calA{{\ensuremath{\cal A}}\xspace}

\def\ra                 {\ensuremath{\rightarrow}\xspace}
\def\to                 {\ensuremath{\rightarrow}\xspace}

\def\pep2{PEP-II}

\def\gsim{{~\raise.15em\hbox{$>$}\kern-.85em
          \lower.35em\hbox{$\sim$}~}\xspace}
\def\lsim{{~\raise.15em\hbox{$<$}\kern-.85em
          \lower.35em\hbox{$\sim$}~}\xspace}

\def\CP                {\ensuremath{C\!P}\xspace}
\def\CPT               {\ensuremath{C\!PT}\xspace} 

\def\T       {\ensuremath{T}\xspace}

\def\deltaz{\ensuremath{{\rm \Delta}z}\xspace}
\def\deltat{\ensuremath{{\rm \Delta}t}\xspace}
\def\deltamd{\ensuremath{{\rm \Delta}m_d}\xspace}

\xspace

\newcommand{\jplBase}        {Phys.\ Lett.\xspace}

\newcommand{\nimBaseC}       {Nucl.\ Instr.\ and Methods\xspace}

\newcommand{\npBase}         {Nucl.\ Phys.\xspace}
\newcommand{\zpBase}         {Z.\ Phys.\xspace}

\newcommand{\nim}       [1]  {\nimBaseC~{\bf #1}}
\newcommand{\npb}       [1]  {\npBase\ B~{\bf #1}}

\newcommand{\plb}       [1]  {\jplBase\ B~{\bf #1}}

\newcommand{\zpc}       [1]  {\zpBase\ C~{\bf #1}}

\def\jetset74   {\mbox{\tt Jetset \hspace{-0.5em}7.\hspace{-0.2em}4}\xspace}

\begin{document}

\title{{\large \bf
Limits on the Decay-Rate Difference of Neutral \boldmath{\B} Mesons and on\\
\boldmath{\CP}, \boldmath{\T}, and \boldmath{\CPT} Violation in
\BzBzb Oscillations
}}

%
\author{B.~Aubert}
\author{R.~Barate}
\author{D.~Boutigny}
\author{J.-M.~Gaillard}
\author{A.~Hicheur}
\author{Y.~Karyotakis}
\author{J.~P.~Lees}
\author{P.~Robbe}
\author{V.~Tisserand}
\author{A.~Zghiche}
\affiliation{Laboratoire de Physique des Particules, F-74941 Annecy-le-Vieux, France }
% ...
\collaboration{The \babar\ Collaboration}
\noaffiliation

\date{\today}

\begin{abstract}
Using events in which one of two neutral \B\ mesons
from the decay of an \FourS\ meson is fully reconstructed,
we determine parameters governing decay (\dGoG), 
\CP and \T violation (\absqop), and \CP and \CPT violation
\mbox{(\reZ, \imZ)}. The results, obtained from an analysis of
88 million $\Upsilon(4S)$ decays recorded by \babar, are
$$\begin{array}{
r@{\ \ =\ }r@{.}l@{\ \pm 0.}l@{{\ \rm{(stat.)}}\ \pm 0.}
l@{{\ \rm{(syst.)}}\ \ [\ }r@{\ ,\ }l@{\ ]}l}
\sgndGoG&-0&008&037&018&-0.084&0.068&~,\\
\absqop     &1 &029&013&011& 1.001&1.057&~,\\
\reZparflat &0 &014&035&034&-0.072&0.101&~,\\
\imZ        &0 &038&029&025&-0.028&0.104&~.
\end{array}$$
The values inside square brackets indicate the 90\% confidence-level
intervals. These results are consistent
with Standard Model expectations.
\end{abstract}

\pacs{13.25.Hw, 12.15.Ff, 11.30.Er}

\maketitle

In this Letter, we provide a direct limit on the total decay-rate
difference \dG between the \Bd mass eigenstates, and set limits on
\CP, \T, and \CPT violation inherent in the mixing of neutral \B
mesons. In the Standard Model \CPT violation is forbidden and the
other effects are expected to be non-zero but small, but new physics
could provide enhancements~\cite{SM-dG-calc,SM-qp-calc,cpt-bosc1,cpt-bb}.  We
test these predictions by analyzing the time dependence of decays of
the \FourS\ resonance in which one neutral \B\ meson (\Brec) is fully
reconstructed and \ed{the flavor of} the other \B\ (\Btag) is identified as being either \Bz\ or \Bzb.
\ed{The \Brec sample is composed of flavor- and
  \CP-eigenstate subsamples, \Bflav and \BCP.}
We reconstruct the flavor \ed{eigen}states~\cite{charge-conj}
\Bflav= $D^{(*)-}\pip(\rho^+,a_1^+)$ and
$\jpsi\Kstarz(\rightarrow\Kp\pim)$ and the \CP eigenstates
\BCP=$\jpsi\KS$, $\psi(2S)\KS$, $\chi_{c1}\KS$, and $\jpsi\KL$.  The
\ed{flavor of the}
\B\ that is not completely reconstructed is ``tagged''
on the basis of the charges of leptons and kaons, and other
indicators~\cite{long-paper}.
The data come from 88 million
$\FourS \to \B\Bb$ decays collected with the \babar\
detector~\cite{babar-detector} at the \pep2\ asymmetric-energy \B
Factory at SLAC.

The light and heavy \Bd mass eigenstates $B_{L,H}$ are 
superpositions of \Bz and \Bzb. This mixing is a consequence of
transitions between \Bz and \Bzb through
intermediate states. \ed{Flavor} oscillations between \Bz and \Bzb occur with a
frequency $\dM\equiv m_H - m_L$. A state that is initially \Bz (\Bzb) will
develop a \Bzb (\Bz) component over time, whose amplitude is
proportional to a complex factor denoted \qop (\poq)~\cite{pdg2002}.
Since $\absqop\simeq 1$ in the Standard Model, this
factor is usually assumed to be a pure phase.

The most general time dependence allowed for the decays of the 
two neutral \B\ mesons coming from an \FourS is~\cite{long-paper}
\begin{multline}
\frac{{\rm d}N}{{\rm d}\deltat} \propto
e^{-\G |\deltat|}\;\biggl[
\frac{|a_{+}|^2 + |a_{-}|^2}{2}\cosh\left(\frac{\dG\deltat}{2}\right)\,+\\
\frac{|a_{+}|^2 - |a_{-}|^2}{2}\cos(\dM\deltat) -
\re(a_{+}^\ast a_{-})\sinh\left(\frac{\dG\deltat}{2}\right)\\
+\,\im(a_{+}^\ast a_{-})\sin(\dM\deltat)
\biggr]\; ,
\label{eq:timedist}
\end{multline}
where $\deltat\ed{\equiv t_{\rmrec}-t_{\rmtag}}$ is the \ed{signed} difference in proper decay times, \G is the
mean decay rate of the two neutral mass eigenstates, and
$\dG\equiv\Gamma_H-\Gamma_L$ is their decay-rate difference.  The
values of the complex parameters $a_\pm$ differ for the various
combinations of flavor and \CP eigenstates into which the \B\ mesons
decay~\cite{long-paper}.

In the simplest picture, where $\dG=0$, and \CP, \T, and \CPT
violation in mixing are neglected, if the fully reconstructed state is
a flavor eigenstate the time distributions
\ed{${{\rm d}N}/{{\rm d}\deltat}$} with perfect tagging are
proportional to $e^{-\G |\deltat|}(1\pm\cos(\dM\deltat))$.  In
practice, the tagging is imperfect and its performance is measured
directly from the data.  Imperfect tagging reduces the coefficient of
$\cos(\dM\deltat)$ by a factor $1-2w$ called the dilution, where $w$
is the probability of tagging incorrectly.

\B decays to a \CP eigenstate $f_{\CP}$ are conveniently parameterized
by $\lCP\equiv (q/p)\,({\ocalA}_{\CP}/{\calA}_{\CP})$, where
$\calA_{\CP}$ (${\ocalA}_{\CP}$) is the amplitude for $\Bz\to f_{CP}$
($\Bzb\to f_{CP}$). \CP violation is characterized by $\lCP\ne
\eta_{\CP}$ where $\eta_{\CP}= \pm 1$ is the final state's \CP
eigenvalue.  The \CP violation observed in decays like $\B\to
\jpsi\KS$~\cite{babar-sin2b,belle-sin2b} involves interference between
decays with and without net oscillation, and leads to $\imlCP\ne 0$.
Other possible sources of \CP violation are $\absqop\ne 1$ and
$|\ocalA_{\CP}/\calA_{\CP}|\ne 1$. We include a test of the former
possibility here.

The time distributions \ed{${{\rm d}N}/{{\rm d}\deltat}$} for the \BCP
samples, in the simplest picture \ed{(defined above)}
and with perfect tagging, are proportional to
\begin{multline}
e^{-\Gamma_d |\deltat|}\,\bigl[1+|\lCP|^2
  \pm(1-|\lCP|^2)\cos(\dM\deltat)
\\
\mp 2\im\lCP\sin(\dM\deltat)\bigr]
\; .
\end{multline}
In the Standard Model we have $\lCP=-e^{-2i\beta}$ for $\jpsi\KS$ with
the approximation $\dG=0$, where $\beta \equiv \rm arg \left[
-V^{}_{\rm cd} V^\ast_{\rm cb} / V^{}_{\rm td} V^\ast_{\rm tb}
\right]$ is one of the angles of the triangle~\cite{beta} that
represents the unitarity of the quark mixing matrix $\rm V^{}_{ij}$.  Since
$|\lCP|=1$, the $\cos(\dM\deltat)$ term is absent.  Again, wrongly
tagged events reduce the amplitude of the oscillatory terms.

To measure \dG, or \CP, \T, or \CPT violation in mixing alone we need
to find small deviations from these simple patterns. Other effects
that can mimic the behavior we seek must be included in the
analysis. Among these are asymmetries in the response of the
detector to \Bz and \Bzb decays~\cite{long-paper} and interference between
dominant and suppressed decay amplitudes to flavor eigenstates, both
those that are fully reconstructed and those that contribute to
tagging~\cite{dcsd-paper,long-paper}.

The time dependence of the \BCP sample includes a
\ed{$\sinh(\dG\deltat/2)$} term that is effectively linear
in \dG, while the flavor sample has an effective second-order sensitivity to
\dG \ed{through a $\cosh(\dG\deltat/2)$ term}. Untagged data are
included in this analysis and improve our
sensitivity to \dG since the contributions of \dG-dependent terms do
not depend on whether \Btag is a \Bz or \Bzb. With our sample
sizes and small measured value of \dG, the \BCP sample dominates our
determination of \dGoG.  While \dM has been well measured
previously~\cite{babar-dm,belle-dm-cpt,belle-dm}, there is only a weak
limit, \ed{$|\dG|/\G<0.18$ at 95\% CL~\cite{delphiDG}}, on \dG.  A recent
theoretical calculation gives $\dG/\G=-0.003$~\cite{SM-dG-calc}.

Violation of \CP and \T in mixing leads to a difference between the
$\Bz\ra\Bzb$ and $\Bzb\ra\Bz$ transition rates proportional to
$|q/p|^4-1$. Our sensitivity to $|q/p|$ comes mostly from the large
flavor-eigenstate sample.  \ed{Previous measurements,
  obtained assuming $\dG=0$,
  give $\absqop-1 = (-0.7\pm 6.4)\times 10^{-3}$~\cite{pdg2003}.}
The Standard Model expectation is
$\absqop-1 = (2.5-6.5)\times 10^{-4}$~\cite{SM-qp-calc}.
 
\CPT violation in mixing enters the time dependence through the
complex quantity
\begin{eqnarray}
\z & \equiv & \frac{\delta m_d-\frac i2\,\delta\Gamma_d}{ \dM - \frac
  i2\dG}~,\label{eq:z-defn}
\end{eqnarray}
where $\delta m_{d}$ ($\delta\Gamma_{d}$) is the $\Bz-\Bzb$ difference
of effective mass (decay rate) expectation values for the \Bz and \Bzb
flavor eigenstates. A non-zero value of either $\delta m_{d}$ or
$\delta\Gamma_{d}$ is only possible if both \CP and \CPT are violated.
The dominant contribution of \imZ to the time dependence is through
the coefficient of $\sin(\dM\deltat)$ for flavor eigenstates, while
\reZ contributes primarily to the coefficients of
$\cosh(\dG\deltat/2)\approx 1$ and $\cos(\dM\deltat)$ for \CP
eigenstates.  The measurement of \z presented here is more general
than previous analyses based on \B\
decays\ed{, which obtained $\imZ= 0.040\pm 0.032\pm
  0.012$~\cite{opal-cpt}, and $\reZ= 0.00\pm 0.12\pm 0.02$, $\imZ=
  -0.03\pm 0.01\pm 0.03$~\cite{belle-dm-cpt},} and complements earlier
limits on 
the $K^0-\Kbar^0$ mass difference $\delta m_K/m_K <
10^{-18}$~\cite{pdg2002}.

Interference effects between the amplitudes for dominant decays of
flavor eigenstates (e.g., $\Bz\to D^-\pi^+$) and for
doubly-CKM-suppressed (DCS) decays (e.g., $\Bzb\to D^-\pi^+$) are
analogous to the interference familiar in decays to \CP
eigenstates~\cite{dcsd-paper}. They thus affect, in particular, the
$\sin(\dM\deltat)$ terms and have the potential to obscure a similar
contribution from \imZ.  The size of the DCS interference relative to
the dominant \ed{\Bz} decay is governed by \ed{
  $\lambda_{Bf}$ and $\lambda_{Bt}$, for \Bflav and \Btag states, respectively.
These parameters are} defined analogously to $\lambda_{\CP}$, and we expect
$|\ed{\lambda_{Bf,Bt}}|\approx |q/p| |V^{}_{\rm ub}V^{\ast}_{\rm
cd}/V^\ast_{\rm cb}V^{}_{\rm ud}|\simeq 0.02
|q/p|$~\ed{\cite{long-paper}}. There are similar
interference contributions from DCS amplitudes for \ed{\Bzb} decays,
\ed{governed by $\lambda_{\overline{B}f}$ and $\lambda_{\overline{B}t}$.}
We write
${\overline\lambda}_{\ed{\overline{B}f,\overline{B}t}}=
1/\lambda_{\ed{\overline{B}f,\overline{B}t}}$ 
so $|{\overline\lambda}_{\ed{\overline{B}f,\overline{B}t}}|
\approx 0.02 |p/q|$.
\ed{The \Bflav and \Btag samples are ensembles of final states that each
contribute to the expected decay-rate distributions with different
amplitudes. We find that, working to first 
order in the small quantities $|\lambda_{Bf,Bt}|$,
$|\overline{\lambda}_{\overline{B}f,\overline{B}t}|$, $|\z|$, and
$\absqop-1$, the cumulative effect of each ensemble does not modify
the expected decay-rate distributions, once $\lambda_{Bf,Bt}$ and
$\overline{\lambda}_{\overline{B}f,\overline{B}t}$ are reinterpreted
as effective parameters.}

We combine all the data for the \CP eigenstates, taking into
account the \CP eigenvalue of the final state.  We assume
$|{\ocalA}_{\CP}/\calA_{\CP}|=1$ (but vary this ratio as a
systematic study) as expected theoretically at the $10^{-3}$
level~\cite{SM-A-calc} and as supported by the average of \B-Factory
measurements of states of charmonium and \KS or \KL, for which \ed{it was}
found $|\ocalA_{\CP}/\calA_{\CP}| = 0.949\pm
0.045~$~\cite{babar-sin2b,belle-sin2b}, when $\dG=0$
and $\absqop=1$ are assumed.

The time interval $\deltat$ between the
two $B$ decays is calculated from the measured separation \deltaz
between the decay vertices of \Brec and \Btag along the collision
axis~\cite{babar-sin2b}.  We determine the position of the
\Brec vertex from its charged tracks. The \Btag decay vertex is
determined by fitting to a common vertex tracks not belonging to the
\Brec candidate, employing constraints from the beam spot location and
the \Brec momentum~\cite{babar-sin2b}.  The r.m.s. \deltat resolution
for 99.7\% of the events used is 1.0\ps, \ed{to be compared with
  $\langle|\deltat|\rangle \simeq 1.5$\ps.}

Resolution effects for signal events are modeled by convolving the
idealized decay rate with a sum of three Gaussian distributions, two
of whose widths and biases are scaled with each event's estimated
\deltat uncertainty $\sigma_{\deltat}$.

We use four mutually-exclusive categories to assign tags, based on
kinematic, particle type, and charge information~\cite{babar-sin2b}.
There are separate reconstruction efficiencies and mistag
probabilities for \Bz and \Bzb tags, to accommodate differences in
detector response to \Bz and \Bzb decays.  In addition, we introduce a
linear dependence of the mistag probability on $\sigma_{\deltat}$,
except for events tagged with a high-momentum lepton.

Backgrounds are primarily due to misreconstructed \Brec candidates and
are studied in data using mass or energy sidebands.  Events are
assigned signal and background probabilities based on their proximity
to the signal peak. We model backgrounds with empirical \deltat
distributions that can accommodate contributions from decays with a
range of lifetimes.

The parameters of primary interest in this analysis are \sgndGoG,
\absqop, \reZparflat, and \imZ. We cannot determine \dGoG and \reZ
directly because both occur multiplied by $\re\lCP$ in their dominant
contributions to the decay rate. They are thus subject to a sign
ambiguity, which can be resolved by relying on additional information
from the unitarity triangle.  The average lifetime $\tauB\equiv 1/\G$
is fixed at $1.542$~ps~\cite{pdg2002}.  The parameters \dM and
\imlambcpflat are determined together with the main parameters as
cross checks against earlier measurements~\cite{babar-dm,
babar-sin2b}.  The terms proportional to the real parts of the
\ed{effective} DCS
parameters \ed{$\lambda_{Bf,Bt}$ and
${\overline{\lambda}_{\overline{B}f,\overline{B}t}}$},
are small and
therefore neglected in the nominal fit model, \ed{while the imaginary
  parts and magnitudes of these effective parameters are} 
treated as independent variables.  For all sets of nonleptonic flavor
eigenstates analyzed, the magnitude of each $|\lambda|$ is fixed to
0.02 (up to a factor $|q/p|$ or $|p/q|$) but $\im\lambda/|\lambda|$ is
left unconstrained.  The decay model uses 26 more parameters to model
the effects of experimental \deltat\ resolution (10), $\Bz/\Bzb$
tagging capability (11), and reconstruction and tagging efficiencies
(5). An additional 22 parameters model the levels and \deltat
dependence of backgrounds.  A total of 58 free parameters are
determined with a simultaneous unbinned maximum-likelihood fit to the
\deltat distributions of \CP and flavor-eigenstate
samples~\cite{long-paper}.

Table~\ref{tab:results} summarizes the results of fits allowing (\z
free) or not allowing ($\z= 0$) \CPT violation in $\Bz\Bzb$
oscillations. The largest statistical correlations
involving the parameters of interest are between \absqop and
parameters modeling $\Bz\Bzb$ asymmetries in reconstruction
efficiency and mistag probabilities, and between \imZ and the 
DCS contributions to \Btag decay amplitudes. The fitted values of \dM
and \imlambcpflat are consistent with recent \B-Factory
measurements~\cite{babar-dm,belle-dm,babar-sin2b,belle-sin2b}. 
When \z is fixed, the value of \imlambcpflat decreases by $0.011$,
equal to 15\% of the statistical uncertainty on \imlambcpflat which is
consistent with the correlations observed in the fit with \z free,
while the value of and uncertainty in \dM are
unchanged. No statistically significant \Bz-\Bzb differences in
reconstruction and tagging efficiencies are observed.

\begin{table}
\caption{\label{tab:results}Fit results allowing (\z free) or not
allowing ($\z= 0$) \CPT violation in \BzBzb oscillations.}
\begin{ruledtabular}
\begin{tabular}{lcc}
Parameter & \z Free & $\z= 0$ \\
\hline
\sgndGoG    & $-0.008\pm 0.037$ & $-0.009\pm 0.037$ \\
\absqop     & $ 1.029\pm 0.013$ & $ 1.029\pm 0.013$ \\
\reZparflat & $ 0.014\pm 0.035$ & \textemdash \\
\imZ        & $ 0.038\pm 0.029$ & \textemdash \\
\end{tabular}
\end{ruledtabular}
\end{table}

We have used data and Monte Carlo samples to validate our analysis
technique. Tests with large, parameterized Monte Carlo samples
demonstrate that the observed statistical uncertainties and correlations are
consistent with expectations. Analyses of Monte Carlo samples
generated with a detailed detector simulation verify that the analysis
procedure is unbiased. Fits to data subsamples selected by tagging
category, running period, and \Brec decay mode give consistent
results. Changes to the algorithms used to estimate \deltat and
$\sigma_{\deltat}$ or to their allowed ranges also have no
statistically significant effect. Fits to samples of charged
\B decays, in which no oscillations are present, give the expected results.

We identify four general sources of systematic uncertainty with the
contributions shown in Table~\ref{tab:syst} for the fit in which \z
is free~\cite{long-paper}. The first is
possible bias in the event selection
and fit method: we see no evidence of such bias when analyzing Monte
Carlo samples and assign the statistical uncertainty of these checks
as a systematic uncertainty on the final results. The second is
the \deltat measurement. The choice of
parameterization of the resolution function dominates this
uncertainty, but assumptions about the beam spot and detector
alignment contribute as well. Assumptions about the properties of
signal $\Upsilon(4S)\rightarrow\Brec\Btag$ decays
include the values of the lifetime, $|\ocalA_{\CP}/{\calA}_{\CP}|$,
and DCS parameters, and are the third source of systematic uncertainty.
Uncertainties in the size and \deltat distributions
of background events (BG) incorrectly identified as
$\Upsilon(4S)\rightarrow\Brec\Btag$ make small contributions to the
systematic uncertainties.

\begin{table}
\caption{\label{tab:syst}Summary of systematic uncertainties (\z free).}
\begin{ruledtabular}
\begin{tabular}{lcccc}
       & $\sign(\relCP)$ & & $\relCP/|\lCP|$ & \\
Source & $\times\dGoG$ & \absqop & $\times\reZ$ & \imZ \\
\hline
Analysis Method       & 0.006 & 0.007 & 0.005 & 0.016 \\
\deltat Resolution    & 0.013 & 0.003 & 0.008 & 0.016 \\
Signal Properties     & 0.010 & 0.008 & 0.033 & 0.009 \\
BG Properties         & 0.005 & 0.003 & 0.007 & 0.004 \\
\hline
Total                 & 0.018 & 0.011 & 0.034 & 0.025 \\
\end{tabular}
\end{ruledtabular}
\end{table}

Different sources dominate the systematic uncertainty for each
parameter. Most systematic uncertainties are determined with data and will
decrease with additional statistics. The largest single source of
uncertainty is the contribution of the DCS parameters to \reZparflat
and it is estimated by varying the DCS phase parameters over their
full allowed range, and $|\ocalA_{Bf}/\calA_{Bf}|$ and
$|\calA_{\overline Bf}/\ocalA_{\overline Bf}|$ over the range 0--$0.04$.
Systematic uncertainties on \sgndGoG and \absqop for the analysis assuming
$\z= 0$ were evaluated similarly as $\pm0.018$ and $\pm0.011$,
respectively.

\begin{figure}[htb]
\includegraphics[width=\columnwidth]{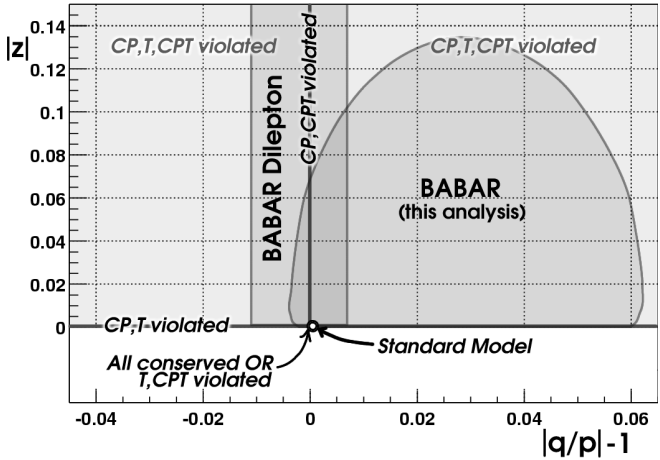}
\caption{\label{fig:symmetries} Favored regions at 68\% confidence
level in the $(\absqop-1,|\z|)$ plane determined by this analysis
and by the \babar\ measurement of the dilepton
asymmetry~\cite{babar-dilepton}. Labels reflect the
requirements that both \CP and \T\
be violated if $\absqop\ne 1$ and that both \CP and \CPT be violated
if $|\z|\ne 0$. The dilepton measurement constrains $\absqop$ without
assumptions on the value of $|\z|$. The Standard Model expectation of
$\absqop-1 = (2.5-6.5)\times 10^{-4}$ is
obtained from Ref.~\cite{SM-qp-calc}.}
\end{figure}

Using the world-average value of \dM~\cite{pdg2002}, we derive the
value $\sign(\relCP)\dG/\dM = -0.011\pm 0.049$(stat.) $\pm
0.024$(syst.), corresponding to the range [-0.11\ed{2},0.091] at the 90\%
confidence level, from the fit results with \z free.  The limit on
\CP and \T violation in oscillations is independent of and consistent
with our previous measurement based on an analysis of inclusive
dilepton events~\cite{babar-dilepton}.  Using Eq.~(\ref{eq:z-defn})
and taking the world-average \Bd mass~\cite{pdg2002}, we derive
$|\delta m_{d}|/m_{\Bd} < 1.0\times 10^{-14}$ and $-0.156 <
\delta\Gamma_{d}/\Gamma_{d} < 0.042$ at the 90\% confidence level.
Figure~\ref{fig:symmetries} shows the results of the fit with \z free
in the $(\absqop-1,|\z|)$ plane, compared to the previous \babar\
measurement of \absqop, and to Standard Model expectations.

Conventional analyses of oscillations and \CP
violation in the \Bd system neglect possible contributions from
several sources that are expected to be small in the Standard
Model. This analysis includes these effects and finds results
consistent with Standard Model expectations.  While the Standard Model
predictions for \dG, \absqop, and \z are well below our current
sensitivity, higher-precision measurements may still bring surprises.

We are grateful for the excellent luminosity and machine conditions
provided by our \pep2\ colleagues, 
and for the substantial dedicated effort from
the computing organizations that support \babar.
The collaborating institutions wish to thank 
SLAC for its support and kind hospitality. 
This work is supported by
DOE
and NSF (USA),
NSERC (Canada),
IHEP (China),
CEA and
CNRS-IN2P3
(France),
BMBF and DFG
(Germany),
INFN (Italy),
FOM (The Netherlands),
NFR (Norway),
MIST (Russia), and
PPARC (United Kingdom). 
Individuals have received support from the 
A.~P.~Sloan Foundation, 
Research Corporation,
and Alexander von Humboldt Foundation.


\begin{thebibliography}{99}
\bibitem{charge-conj}
Charge conjugation is implied throughout this letter, unless
explicitly stated otherwise.
\end{thebibliography}
\end{document}